\documentclass[lettersize,journal]{IEEEtran}

\usepackage[linesnumbered,ruled,vlined]{algorithm2e}
\usepackage{cite} 
\usepackage{amsmath} 
\usepackage{array,color}
\usepackage{xstring}
\usepackage{textcomp}
\usepackage{stfloats}
\usepackage{url}
\usepackage{verbatim}
\usepackage{graphicx}
\usepackage{orcidlink}
\usepackage{nomencl}
\usepackage{caption}
\usepackage{subcaption}
\usepackage[utf8]{inputenc}
\usepackage{pdfpages}
\usepackage{xcolor}
\usepackage{float}

\renewcommand\nomgroup[1]{%
  \item[\bfseries
  \ifstrequal{#1}{A}{Acronyms}{%
  \ifstrequal{#1}{M}{Mathematical Symbols}{}}%
]}

\fontsize{9.96}{11.96}\selectfont 

\usepackage[letterpaper, left=0.673in, right=0.673in, top=0.767in, bottom=0.607in]{geometry}

\def\BibTeX{{\rm B\kern-.05em{\sc i\kern-.025em b}\kern-.08em
    T\kern-.1667em\lower.7ex\hbox{E}\kern-.125emX}}

\begin{document}

\title{AI-Driven Secure Data Sharing: A Trustworthy and Privacy-Preserving Approach}

\author{\IEEEauthorblockN{Al Amin\IEEEauthorrefmark{1}, Kamrul Hasan\IEEEauthorrefmark{1}, Sharif Ullah\IEEEauthorrefmark{2}, Liang Hong\IEEEauthorrefmark{1}}\\
    \IEEEauthorblockA{\IEEEauthorrefmark{1}Tennessee State University, TN, USA}\\
    \IEEEauthorblockA{\IEEEauthorrefmark{2}University of Central Arkansas, Conway, AR, USA}\\
    Email: \texttt{\{aamin2, mhasan1, lhong1\}@tnstate.edu}, \texttt{mullah@uca.edu}
}

\maketitle

\begin{abstract}

In the era of data-driven decision-making, ensuring the privacy and security of shared data is paramount across various domains. Applying existing deep neural networks (DNNs) to encrypted data is critical and often compromises performance, security, and computational overhead. To address these limitations, this research introduces a secure framework consisting of a learnable encryption method based on the block-pixel operation to encrypt the data and subsequently integrate it with the Vision Transformer (ViT). The proposed framework ensures data privacy and security by creating unique scrambling patterns per key, providing robust performance against adversarial attacks without compromising computational efficiency and data integrity. The framework was tested on sensitive medical datasets to validate its efficacy, proving its capability to handle highly confidential information securely. The suggested framework was validated with a 94\% success rate after extensive testing on real-world datasets, such as MRI brain tumors and histological scans of lung and colon cancers. Additionally, the framework was tested under diverse adversarial attempts against secure data sharing with optimum performance and demonstrated its effectiveness in various threat scenarios. These comprehensive analyses underscore its robustness, making it a trustworthy solution for secure data sharing in critical applications.

\end{abstract}

\begin{IEEEkeywords}
Privacy-Preserving Data Sharing, Vision Transformer (ViT), Adversarial Robustness, Data Integrity, and Confidentiality.
\end{IEEEkeywords}

\section{Introduction}

The critical importance of privacy-preserving data sharing, which ensures both the integrity and confidentiality of sensitive information, is paramount across diverse sectors \cite{7440648}, protecting against unauthorized access and maintaining trust in data exchanges. Concurrently, scholars have increasingly emphasized machine learning (ML) and deep learning (DL) architectures to ensure data security and improve model performance as well as reduce computational overhead \cite{10273222}. As a result, cloud-based platforms such as Google Cloud and Microsoft Azure have recently become popular because DL architecture requires a large amount of data for accurate diagnosis. Therefore, a cloud-based platform is suitable for reducing computational overhead and model training time \cite{DBLP:journals/corr/abs-2001-07761}. However, it poses a potential risk of privacy and security breaches due to the broad use of these shared cloud environments. Addressing these privacy and security challenges has thus become an urgent priority.

Privacy-preserving data sharing is vital across diverse applications, safeguarding sensitive information while ensuring the integrity and confidentiality of data exchanges \cite{10074197}. As industries from finance to education increasingly rely on data-driven decisions, the need for robust privacy measures becomes more critical \cite{10398775}. In this research, medical imaging serves as a primary use case, illustrating the specialized requirements for privacy in healthcare applications. The medical domain demands stringent privacy controls due to the sensitive nature of health data, where breaches can have profound personal and legal repercussions. The Health Insurance Portability and Accountability Act (HIPAA) Privacy Rule allows individuals to see their health records and sets national requirements to keep them safe. A safe, patient-centered healthcare system will enable people to track their health, follow their treatment programs, and participate in research \cite{Health:Privacy, HIPPA:Right}. Researchers have recently focused on the challenge of classifying encrypted images using deep neural networks (DNNs) to mitigate these security concerns\cite{kiya2023blockwiseencryptionreliablevision,9802995}. Traditional perceptual encryption methods protect image details and embed unique features controlled by a key; however, they often degrade the performance of DNNs due to the added complexity of processing encrypted data, which affects model accuracy and efficiency. 

To address the inherent challenges of secure and trustworthy data sharing across various domains, the proposed framework introduces a learnable encryption method that enhances robustness while maintaining high performance with reduced computational overhead compared to existing approaches. The goal of this encryption method integrated into the proposed approach is to securely obfuscate the data while preserving essential features necessary for accurate classification. We rigorously tested this framework by evaluating its performance across different medical datasets in an encrypted format, including MRI brain tumors and lung and colon cancer histopathological images. This achieved a validation accuracy of 94\% with considerable processing time. To validate the effectiveness of the framework against adversarial attempts, we evaluate our approach through diverse attacks. The proposed learnable encryption method was tested with data reconstruction attacks such as Leading Bit attacks and Minimum Difference attacks, where our method demonstrates strong protection of sensitive medical data in cloud-based artificial intelligence (AI) services. Furthermore, to assess it's robustness on integrity attacks, Bit-flip and Gaussian Noise attacks are applied by poisoning 10\% and 20\% of the encrypted training data before the training model.  Despite these adversarial perturbations, the proposed framework maintained classification accuracy between 90\% and 85\%, demonstrating its resilience in protecting medical data without compromising performance. In summary, the main contributions of this work are as follows:

\begin{itemize}
    \item This research introduces a framework that integrates Vision Transformer (ViT) with \emph{learnable encryption based on Block-Pixel operation,} creating unique scrambling patterns per key, ensuring robust security and privacy in medical image sharing and classification.
    
    \item We demonstrated the significance of the proposed approach within diverse threat scenarios: data reconstruction attack and integrity attack,  affirming the robust data security in AI services in cloud environments. 

    \item  The proposed framework achieved a 94\% validation accuracy on real-world datasets, including MRI brain tumors and lung and colon cancer images, showcasing its effectiveness in encrypted medical image classification.

\end{itemize}

The rest of the paper is structured as follows: Section II reviews recent research and challenges regarding secure data sharing as well as encryption in medical image analysis. Section III introduces the proposed framework applied to medical image analysis. Section IV details the datasets and experimental results. Section V concludes with key findings and future research directions.

\section{Related Works}

In this section, we explore different security concerns associated with data sharing in AI services. Additionally, we also analyze the existing challenges associated with ML techniques in medical image analysis regarding optimum privacy and accuracy with reduced computational overhead as this domain is selected for our framework's evaluation in this paper. 

\subsection{Secure Data Sharing}
As AI and ML technologies expand across industries, ensuring secure and privacy-preserving data sharing becomes crucial for protecting sensitive information while enabling analytics. Shubyn et al. address this by proposing a federated learning (FL) approach for 5G networks, which enhances data privacy through local processing instead of centralized servers, demonstrating FL’s potential to reduce redundancy while maintaining network efficiency \cite{9767080}. Similarly, Cunha et al. introduce the 5Growth project, which uses AI/ML for secure data sharing in 5G network slices, focusing on data isolation and anomaly detection. Despite these advancements, challenges remain in managing AI/ML complexity and maintaining consistent performance across services \cite{9482536}. In the financial domain, Wang and Tsai apply differential privacy to secure data sharing, illustrating its effectiveness in protecting sensitive information while balancing privacy with model accuracy. Their work highlights the trade-offs faced when incorporating noise to ensure privacy \cite{wang2022protection}. Extending these efforts to mobile healthcare, Huang et al. propose dsPPS, a privacy-preserving framework combining biometric data collection and attribute-based access control for Personal Health Records (PHRs). Although effective, further optimization and verification are required to enhance its efficiency and security \cite{8359554}.

\subsection{Privacy-Preserving Medical Image Classification}
Healthcare industries worldwide are becoming increasingly concerned with patient data privacy \cite{10461067}. Existing research exhibits various encryption systems integrating ML and DL architectures to protect data privacy \cite{9153891}. Madonna et al. \cite{madono2020blockwisescrambledimagerecognition} applied an extended learnable encryption (ELE) system and a deep neural network (DNN) model using an adaptation network to recognize scrambled images, achieving an accuracy of 83.59\% on the CIFAR-100 dataset for plain images without an adaptation network. Similarly, Kiya et al. \cite{kiya2023blockwiseencryptionreliablevision} explored block-wise encryption for reliable vision transformer models, which effectively protected image details but faced performance degradation in DNNs due to the encryption process. A notable limitation of this study is the lack of adversarial tests to evaluate the model's robustness, leading to an incomplete assessment of its security.
Sirichotedumrong et al. \cite{8804201} proposed a pixel-based image encryption method combined with an adaptation network to maintain DNN performance while enabling data augmentation in the encrypted domain. However, challenges remain in accurately representing ground truth in the augmented data, which is crucial in medical imaging to avoid incorrect disease localization and potential clinical consequences. Huang et al. \cite{9802995} introduced an enhanced learnable image encryption scheme for privacy-preserving DL on medical images, incorporating statistical smoothing techniques to maintain privacy while allowing encrypted images to train deep neural networks (DNNs). However, a key limitation is the need to focus on reducing computation time for real-life implementation to lower costs and explore different smoothing processes and diverse datasets for broader applicability. Cao et al. proposed a horizontal federated learning system utilizing homomorphic encryption (FLHE) for multi-institutional lung image classification. This approach enables multiple medical institutions to train models while preserving data privacy collaboratively. Although FLHE enhances data security and confidentiality, it faces significant challenges, particularly in computational complexity and reduced accuracy. The existing research highlights the need for robust encryption techniques to ensure data privacy and security without compromising model performance and computational complexity.

\section{Proposed Approach}

\begin{figure}[h]
    \centering
    \includegraphics[width=\linewidth]{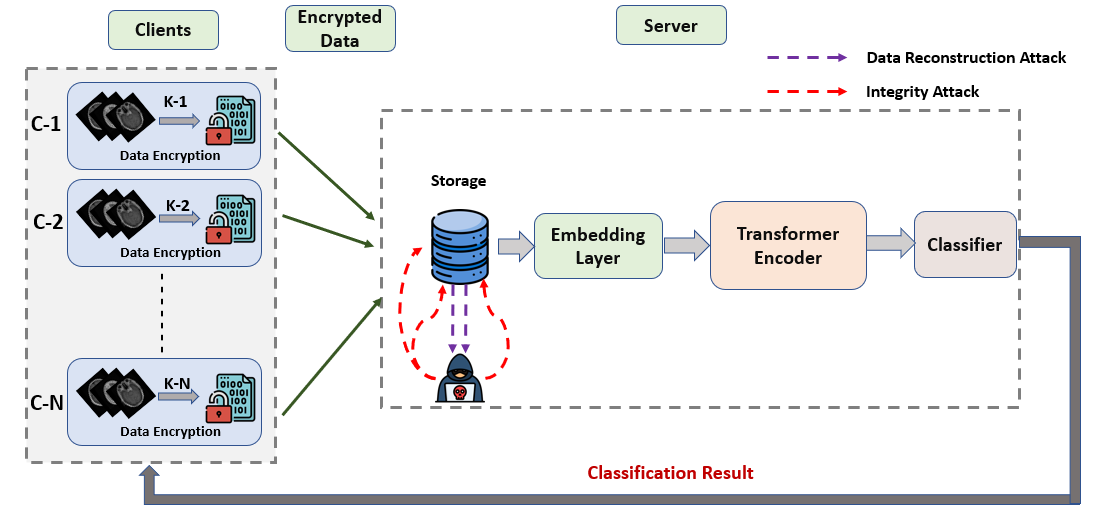}
    \caption{Overview of ViT integrated privacy-preserving secure medical data sharing and classification framework.}
    \label{fig:Pipeline1}
\end{figure}

This research proposes a trustworthy and privacy-preserving framework to ensure secure data sharing and robust classification of medical images. Figure \ref{fig:Pipeline1}, illustrates the implemented approach, where multiple clients (C-1, C-2,..., C-N) encrypt their local data using distinct encryption keys (K-1, K-2,..., K-N). A learnable encryption technique based on the block-pixel operation is applied, incorporating transformations such as negative-positive conversion and color channel shuffling to obfuscate image details while retaining essential features for classification. The encrypted datasets from these clients are transmitted securely to a central server for further processing. At the server, the data passes through an Embedding Layer, which maps the scrambled patches into a high-dimensional space suitable for the Vision Transformer. The Transformer Encoder then processes these high-dimensional representations, capturing intricate dependencies and correlations among the encrypted patches to enable effective feature extraction. A classifier then uses these extracted features to produce accurate classifications. The framework is designed to manage the diverse feature spaces from different data owners, allowing for the construction of an efficient and unified global model. Additionally, it incorporates measures to protect against adversarial attempts, such as data reconstruction and integrity attacks, ensuring the confidentiality and security of sensitive data in cloud-based environments.

\vspace{-5mm}
\subsection{Learnable Encryption based on Block-Pixel Operation}
The proposed Learnable Encryption technique integrates block-wise scrambling, pixel-level transformations, and channel shuffling to secure images effectively. This multi-layered approach obfuscates visual details while preserving essential features required for accurate classification. The encryption process consists of the following components:

\begin{itemize}
    \item \textbf{Image Partitioning into Patches:}
    The input image \( I \) of dimensions \( 200 \times 200 \times 3 \) (Height, Width, Channels) is divided into smaller patches of size \( 25 \times 25 \times 3 \). This operation is represented as:
    \begin{equation}
    I \rightarrow \{P_1, P_2, \dots, P_{64}\}
    \end{equation}
    where \( 64 = \frac{200 \times 200}{25^2} \) is the total number of patches.

    \item \textbf{Block-Wise Scrambling:}
    Each patch \( P_i \) is scrambled using a key \( K \), reordering the pixels within the patch to enhance security. The scrambling function is defined as:
    \begin{equation}
    S(P_i, K) = \text{Permute}(P_i, K)
    \end{equation}
    This process ensures that the pixel arrangement within each patch is randomized in a key-dependent manner, significantly enhancing the security of the encryption process and eliminating recognizable spatial patterns within the image \( S(P_i, K) \) representing the scrambled patch.

    \item \textbf{Shuffling Patch Positions:}
    The scrambled patches are shuffled based on the key \( K \). The original patch positions indexed as \( \{1, 2, \dots, 64\} \) are rearranged as:
    \begin{equation}
    \text{Shuffle}(P_i, K) \rightarrow P_{\sigma(i)}
    \end{equation}
    where \( \sigma \) is a permutation of the indices controlled by the key \( K \) that disrupts the spatial sequence of the scrambled patches, making it highly challenging for adversaries to reconstruct the original spatial arrangement of the image.

    \item \textbf{Negative-Positive Transformation:}
    Pixel values in each patch undergo conditional inversion:
    \begin{equation}
    x' = 
    \begin{cases} 
    x \& \text{if } r = 0 \\
    255 - x & \text{if } r = 1 
    \end{cases}
    \end{equation}
    where \( r \) is a binary value (0 or 1) generated for each pixel based on the key \( K \).

    \item \textbf{Color Channel Shuffling:}
    The RGB channels of each pixel are shuffled according to the key \( K \). If the original channels are \( C = [R, G, B] \), the shuffled channels \( C' \) are:
    \begin{equation}
    C' = [C_{\sigma(1)}, C_{\sigma(2)}, C_{\sigma(3)}]
    \end{equation}
    This step obfuscates the color information of the image by randomly shuffling the RGB color channels based on the key, adding an additional layer of complexity to prevent unauthorized reconstruction.

    \item \textbf{Reconstruction of the Encrypted Image:}
    The final encrypted image \( I_{\text{encrypted}} \) is reconstructed by combining the scrambled and transformed patches:
     \begin{equation}
    \begin{aligned}
       I_{\text{encrypted}} = &\ \text{Reassemble}(\{S(P_1, K), S(P_2, K), \dots, \\
    &\ S(P_{64}, K)\})
    \end{aligned}
    \end{equation}
\end{itemize}

\begin{figure}[ht]
    \centering    \includegraphics[width=\linewidth]{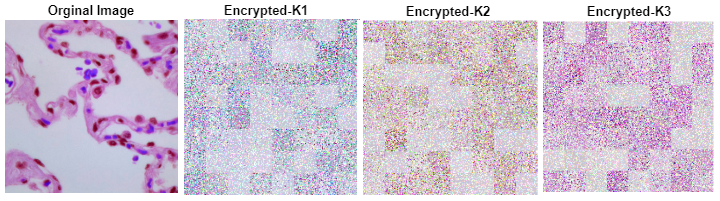}
    \caption{Original image and encrypted images generated using the learnable encryption method based on block-pixel operation with three different keys (\emph{K1, K2, K3}).}
    \label{fig:key}
\end{figure}

This multi-layered encryption technique, combined with unique keys for each client as illustrated in Figure \ref{fig:key}), ensures a robust defense against unauthorized access, making it extremely difficult to revert the image to its original form without the correct key.

\vspace{-4mm}

\subsection{ViT Architecture}

The Vision Transformer (ViT) model is particularly well-suited for tasks requiring capturing complex, long-range dependencies within images, making it ideal for handling encrypted medical images. This research integrated the ViT model with the proposed learnable encryption method to enhance security in medical imaging because ViT directly applies the transformer architecture to sequences of image patches, allowing it to perform well in image classification tasks, even with encrypted data. The following sections describe the multiple modules that comprise this integrated approach.

\subsubsection{Embedding Layer}

The embedding layer in the ViT model transforms the sequence of encrypted image patches into continuous vector representations suitable for processing by the transformer. Each flattened patch is projected into a 768-dimensional embedding space using a learnable embedding matrix \(E\). This transformation encodes spatial features into dense vectors, preserving the patterns within the encrypted data. A classification token \(v_c\) is prepended to the embedded sequence, introducing a global context that integrates local patch details with the overall image structure. The embedding process is formulated as:
\begin{equation}
k_0 = [v_c; \text{Patch}_1E; \text{Patch}_2E; \dots; \text{Patch}_nE] + E_p
\end{equation}
where \(E_p\) denotes positional embeddings that maintain spatial relationships across patches. This configuration provides a rich, context-aware representation, preparing the data for the transformer encoder's processing and subsequent classification.

\subsubsection{Transformer Encoder}

The Transformer Encoder in the Vision Transformer (ViT) model is composed of two primary components: the Multi-Head Self-Attention (MSA) block and the fully connected feedforward neural network, or Multi-Layer Perceptron (MLP). The MSA block is responsible for assessing the relative importance of each encrypted image patch embedding by comparing it with others within the sequence. The MLP then further processes these embeddings, with Layer Normalization (LN) ensuring stable training throughout the layers. The following equations can describe the operations within the Transformer Encoder \cite{dosovitskiy2020image} :
\begin{equation}
z'_{\ell} = \text{MSA}(\text{LN}(z_{\ell-1})) + z_{\ell-1}, \quad \ell = 1, \dots, L
\end{equation}
\begin{equation}
z_{\ell} = \text{MLP}(\text{LN}(z'_{\ell})) + z'_{\ell}, \quad \ell = 1, \dots, L
\end{equation}
For final classification, the output corresponding to the class token from the sequence \( z_{0L} \) is passed through a dense layer with softmax activation.

\subsection{Algorithmic Representation of the Proposed Approach}
This subsection outlines the algorithmic steps of the secure encrypted medical data classification framework (Algorithm \ref{alg:vitest}), emphasizing both data integrity and confidentiality. The framework begins by partitioning each medical image into patches, followed by a sequence of encryption steps: block-wise scrambling, patch shuffling, pixel inversion, and color channel shuffling, all governed by a key. The encrypted patches are reassembled and transformed into high-dimensional vectors through an embedding layer for the Vision Transformer. The Transformer Encoder applies multi-head self-attention and feedforward layers to extract features. The classification head then predicts outcomes based on these embeddings. The framework’s robustness is evaluated through integrity (Bit-Flip and Gaussian Noise) and confidentiality (Leading Bit and Minimum Difference) attacks, ensuring secure and accurate classification.

\begin{algorithm}[ht]
\scriptsize
\caption{Proposed framework for encrypted medical data classification}\label{alg:vitest}
\KwData{Medical Image Dataset}
\KwResult{Encrypted Image Classification with Integrity and Confidentiality Assurance}

\SetKwFunction{FMain}{Main}
\SetKwProg{Fn}{Function}{:}{}
\Fn{\FMain{Medical Image Dataset}}{
    \textbf{Step 1: Data Partitioning}\;
    Divide each image $I$ into $64$ patches of size $25 \times 25 \times 3$\;
    
    \textbf{Step 2: Block-Wise Scrambling}\;
    \For{each patch $P_i$}{
        Apply scrambling $S(P_i, K)$ using key $K$\;
    }
    
    \textbf{Step 3: Shuffling Patch Positions}\;
    Shuffle scrambled patches based on key $K$\;
    
    \textbf{Step 4: Pixel-Level Transformation}\;
    Apply negative-positive inversion to pixels based on key $K$\;
    
    \textbf{Step 5: Color Channel Shuffling}\;
    Shuffle RGB channels of each patch based on key $K$\;
    
    \textbf{Step 6: Image Reconstruction}\;
    Reassemble encrypted patches into complete image $I_{\text{encrypted}}$\;
    
    \textbf{Step 7: Embedding Layer}\;
    Project patches of $I_{\text{encrypted}}$ into $768$-dimensional vectors\;
    
    \textbf{Step 8: Transformer Encoder}\;
    \For{each layer in encoder}{
        Apply multi-head attention and feedforward processing\;
    }
    
    \textbf{Step 9: Classification Head}\;
    Output class predictions using a dense layer with softmax\;
    \textbf{Step 10: Integrity and Confidentiality Testing}\;
    Evaluate performance under \text{Bit-Flip}, Gaussian Noise (integrity), and Leading Bit, Minimum Difference (confidentiality) attacks to verify robustness\;
} 
\end{algorithm}

\subsection{Robustness Evaluation Through Adversarial Attacks}

This section describes the methodology for applying different adversarial attacks to the encrypted images before model training, focusing on evaluating the robustness of the data-sharing approach with model performance.

\subsubsection{Adversarial Noise Injection Bit-Flip and Gaussian Noise Attacks} To assess the framework's robustness against integrity threats, we applied two types of adversarial noise: Bit Flip and Gaussian noise, on the encrypted training data prior to model training. These noise perturbations were systematically applied to simulate real-world imperfections and potential adversarial threats.

\textbf{Bit-Flip Attack}: The bit-flip attack evaluates the model's sensitivity to minute changes in pixel values at the bit level, which could result from hardware errors or malicious interference. The bit-flip operation on an encrypted image \( I \) with dimensions \( H \times W \times C \) (height, width, and channels) is mathematically defined as:
\begin{equation}
    I_{\text{attacked}}(i, j, k) = I(i, j, k) \oplus 2^n
\end{equation}
where \( \oplus \) represents the bitwise XOR operation, \( n \) is the bit position to be flipped, and \( i \), \( j \), and \( k \) index the height, width, and channels of the image, respectively. This introduces pixel-level perturbations to assess the model’s ability to maintain performance despite such noise.

\textbf{Gaussian Noise Attack}: Gaussian noise simulates sensor imperfections or imaging artifacts by introducing random noise into pixel values. It is defined as:
\begin{equation}
    I_{\text{attacked}}(i, j, k) = I(i, j, k) + \mathcal{N}(\mu, \sigma^2)
\end{equation}
where \( \mathcal{N}(\mu, \sigma^2) \) represents Gaussian noise with mean \( \mu \) and variance \( \sigma^2 \). This tests the model’s resilience to noise typically encountered in medical imaging.


\subsection{Reconstruction Attacks: Leading Bit and Minimum Difference Attacks}

The robustness of the proposed encryption scheme lies in its key-dependent transformations; even if an attacker knows the encryption process, the lack of access to the unique encryption key ensures that the pixel-level obfuscation remains highly secure, making image reconstruction infeasible. To evaluate this robustness, we applied two reconstruction attacks: the: \emph{Leading Bit Attack} and \emph{Minimum Difference Attack}. 

\textbf{Leading Bit Attack}: This attack leverages the fact that, in smooth regions of an image, the most significant bits (MSB) of neighboring pixels tend to exhibit similar values. The attacker attempts to exploit this property to reconstruct a recognizable grayscale approximation of the original image from the encrypted data by manipulating the MSB of each pixel \cite{9802995}. 

\textbf{Minimum Difference Attack}: This attack minimizes the pixel differences between neighboring pixels by evaluating color shuffling and negative-positive transformations in our encrypted medical images \cite{9802995}. The goal is to find the pixel transformation that minimizes the color difference:
\begin{equation}
p' = \arg\min_{p' \in \mathcal{P}} \sum_{c \in \{R,G,B\}} | q_c - p'_c |
\end{equation}
It represents pixel transformation \( p' \) that minimizes the sum of absolute differences across color channels (\( R, G, B \)) between the original pixel \( q \) and \( p' \), reducing color variations.

\begin{figure}[t]
    \centering
    \includegraphics[width=\linewidth]{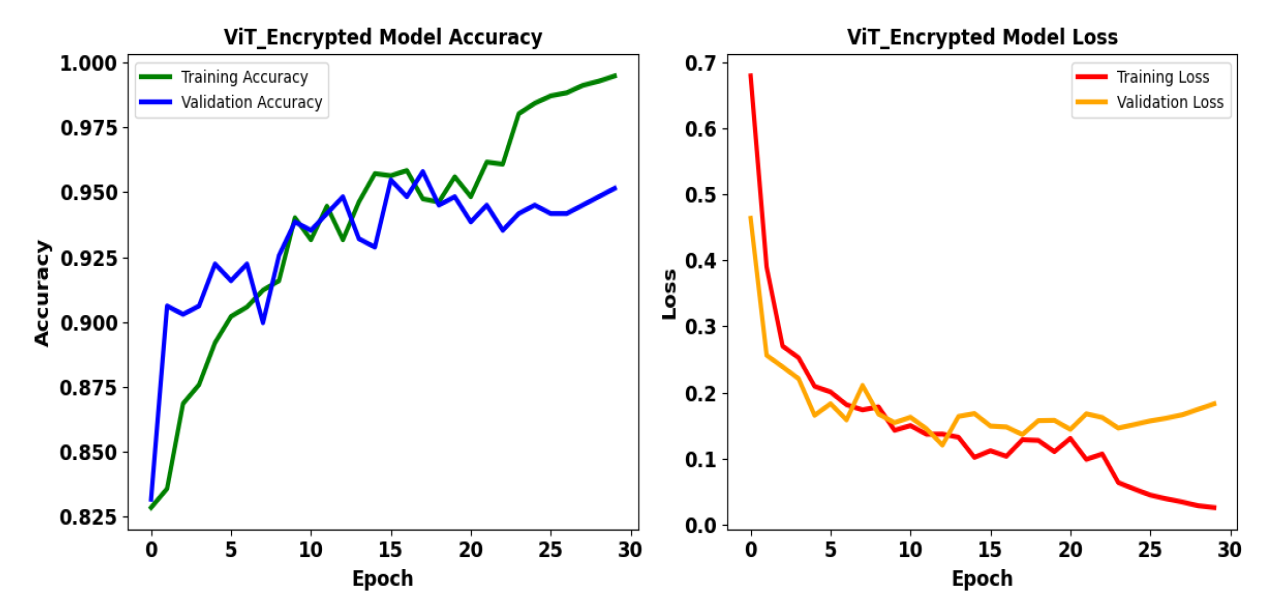}
    \caption{Training and Validation Accuracy and Loss Curves for the proposed Framework.}
    \label{fig:loss_acc}
\end{figure}

\section{RESULTS AND DISCUSSION}

\subsection{Experimental Setup}

The experiments were conducted on a High-Performance Computing (HPC) system equipped with 31 GB of DDR4 RAM and an NVIDIA GeForce GTX 3070 GPU with 8 GB of DDR4 memory, running on Linux Ubuntu. This setup efficiently handled the computational demands of encrypted data processing and Vision Transformer training.

\subsection{Dataset Description}

Two distinct medical image datasets were employed in the investigations \cite{Lung,msoud_nickparvar_2021}. We evaluated the proposed model using diverse datasets, including 5,712 MRI images (with 1,311 used for testing) for brain tumor classification and 2,980 histopathological images (with 311 for testing) for lung and colon cancer classification.

\begin{table}[t]
\centering
\caption{Performance under Adversarial Noise Injection in the proposed framework}
\label{table:noise_training_accuracy}
\begin{tabular}{|l|l|l|l|}
\hline
\textbf{\begin{tabular}[c]{@{}l@{}}Adversarial \\ Attack\\ Type\end{tabular}} &
  \textbf{\begin{tabular}[c]{@{}l@{}}Perturbed \\ Training\\ Dataset\\ (\%)\end{tabular}} &
  \textbf{\begin{tabular}[c]{@{}l@{}}Training\\ Accuracy\\ (\%)\end{tabular}} &
  \textbf{\begin{tabular}[c]{@{}l@{}}Validation \\ Accuracy\\ (\%)\end{tabular}} \\ \hline
\begin{tabular}[c]{@{}l@{}}Gaussian \\ Noise\end{tabular} & 10           & 92          & 90          \\ \hline
\begin{tabular}[c]{@{}l@{}}Gaussian \\ Noise\end{tabular} & 20           & 86          & 85          \\ \hline
Bit-Flip                                                  & 10           & 90          & 88          \\ \hline
Bit-Flip                                                  & 20           & 85          & 84          \\ \hline
\textbf{Without Noise}                                    & \textbf{N/A} & \textbf{95} & \textbf{94} \\ \hline
\end{tabular}
\end{table}

\vspace{-3mm}
\subsection{Model Performance }
The proposed framework significantly outperforms traditional DNN architectures such as InceptionV3, ResNet50, and MobileNetV2 when operating in the encrypted domain. As shown in Table \ref{tab:performance_comparison}, the proposed framework achieves higher training and validation accuracies of 95\% and 94\% on encrypted MRI brain tumor datasets, compared to the DNN models, which exhibit much lower accuracies ranging from 38\% to 51\%. Moreover, the proposed framework demonstrates superior computational efficiency, completing the training process up to 6.58 times faster than the DNN models in the encrypted domain. In contrast, DNN models perform well with plain, unencrypted images, achieving higher accuracies; however, they suffer significant performance degradation when applied to encrypted images. The proposed framework ensures robust performance, data security, and high classification accuracy within the encrypted domain, with training and validation accuracy and loss curves demonstrating steady convergence and improvement, as shown in Figure \ref{fig:loss_acc}.

\begin{table*}[ht]
\centering
\caption{{Performance Comparison of DNN Models and the Proposed Framework on Plain and Encrypted Dataset}}
\label{tab:performance_comparison}
\begin{tabular}{|c|c|c|c|c|c|c|c|}
\hline
\textbf{\begin{tabular}[c]{@{}c@{}}Image \\ Type\end{tabular}} &
  \textbf{Encryption} &
  \textbf{Model} &
  \textbf{Dataset} &
  \textbf{\begin{tabular}[c]{@{}c@{}}Training \\ Acc. (\%)\end{tabular}} &
  \textbf{\begin{tabular}[c]{@{}c@{}}Validation \\ Acc. (\%)\end{tabular}} &
  \textbf{\begin{tabular}[c]{@{}c@{}}Train \\ Time (m)\end{tabular}} &
  \textbf{\begin{tabular}[c]{@{}c@{}}Val. \\ Time (m)\end{tabular}} \\ \hline
{\begin{tabular}[c]{@{}c@{}}Plain \\ Image\end{tabular}} &
  N/A &
  ResNet50 &
  \begin{tabular}[c]{@{}c@{}}Brain Tumor\\ (MRI)\end{tabular} &
  94 &
  93 &
  17.28 &
  4.32 \\ \cline{2-8} 
 &
  N/A &
  MobileNet V2 &
  Brain Tumor (MRI) &
  98 &
  96 &
  31.62 &
  7.9 \\ \hline
{\begin{tabular}[c]{@{}c@{}}Encrypted\\ Image\end{tabular}} &
  \begin{tabular}[c]{@{}c@{}}Pixel \\ Shuffling\end{tabular} &
  ResNet50 &
  Brain Tumor (MRI) &
  46 &
  38 &
  541.6 &
  108.32 \\ \cline{2-8} 
 &
  \begin{tabular}[c]{@{}c@{}}Pixel \\ Shuffling\end{tabular} &
  MobileNet V2 &
  Brain Tumor (MRI) &
  51 &
  51 &
  184.92 &
  46.23 \\ \cline{2-8} 
 &
  \begin{tabular}[c]{@{}c@{}}Pixel \\ Shuffling\end{tabular} &
  Inception V3 &
  Brain Tumor (MRI) &
  48 &
  46 &
  432.36 &
  86.46 \\ \cline{2-8} 
 &
{\textbf{\begin{tabular}[c]{@{}c@{}}Proposed\\ Encryption\end{tabular}}} &
  {\textbf{\begin{tabular}[c]{@{}c@{}}ViT \\ (Proposed)\end{tabular}}} &
  \textbf{Brain Tumor (MRI)} &
  \textbf{95} &
  \textbf{94} &
  \textbf{82.25} &
  \textbf{19.78} \\ \cline{4-8} 
 &
   &
   &
  \begin{tabular}[c]{@{}c@{}}Lung and Colon \\ (Histopathological)\end{tabular} &
  95 &
  94 &
  98.11 &
  24.53 \\ \hline
\end{tabular}
\end{table*}

\begin{figure}[t]
    \centering
    \includegraphics[width=\linewidth]{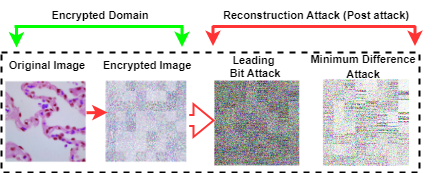}
    \caption{Transformation Pipeline: Original, Encrypted, and Post-Attack Images for Confidentiality Evaluation.}
    \label{fig:attack_image}
\end{figure}

\subsection{Security Evaluation}

The major goal of this work is to ensure the privacy of the data while keeping the optimum AI service. To demonstrate the utility of the approach towards this purpose, we evaluated it's security in terms of both integrity and confidentiality.

\textbf{Integrity Evaluation:}  
To evaluate the framework integrity, Bit-Flip, and Gaussian Noise attacks were applied with 10\% and 20\% perturbations in the training dataset. As shown in Table \ref{table:noise_training_accuracy}, the model achieved 92\% training and 90\% validation accuracy with 10\% Gaussian Noise, and 86\% and 85\% accuracy with 20\% noise. For the Bit-Flip attack, accuracies were 90\% (training) and 88\% (validation) at 10\% perturbation and 85\% and 84\% at 20\%. These results confirm the framework’s robustness in maintaining data integrity despite adversarial noise. It should be noted that in both cases, we assume that attackers knew the encryption method but not the keys chosen by clients. On the other hand, even with full access (i.e., worst case), strategic attackers don't poison the full data as too much perturbation generates more correlated alerts, eventually warning the security administrators with high priority.

\textbf{Confidentiality Evaluation:}  
The framework’s block-wise scrambling and pixel-level transformations effectively resisted Leading Bit and Minimum Difference attacks, preventing the reconstruction of original images and demonstrating its robustness in maintaining the confidentiality of sensitive medical data, as shown in Figure \ref{fig:attack_image}.

\section{Conclusion and Future Work}

This research presents a secure and efficient framework for medical image classification, utilizing a Vision Transformer (ViT) model enhanced with learnable encryption based on Block-Pixel operation. The proposed approach achieves superior accuracy and efficiency compared to traditional DNNs, effectively safeguarding sensitive medical data during cloud-based analysis. Future work will focus on extending this framework to handle multimodal clinical data, such as MRI, CT, and histopathological images, enabling more realistic and comprehensive medical image analysis.

\section{ACKNOWLEDGMENT}
This material is based on work supported by the National
Science Foundation Award Numbers 2205773 and 2219658.

\bibliographystyle{ieeetr}
\bibliography{references.bib}

\begin{thebibliography}{10}

\bibitem{7440648}
M.~Guo, N.~Pissinou, and S.~Iyengar, ``Privacy-aware mobile sensing in vehicular networks,'' in {\em 2016 International Conference on Computing, Networking and Communications (ICNC)}, pp.~1--5, 2016.

\bibitem{10273222}
X.~Zhao, Z.~Qi, S.~Wang, Q.~Wang, X.~Wu, Y.~Mao, and L.~Zhang, ``Rcps: Rectified contrastive pseudo supervision for semi-supervised medical image segmentation,'' {\em IEEE Journal of Biomedical and Health Informatics}, vol.~28, no.~1, pp.~251--261, 2024.

\bibitem{DBLP:journals/corr/abs-2001-07761}
K.~Madono, M.~Tanaka, M.~Onishi, and T.~Ogawa, ``Block-wise scrambled image recognition using adaptation network,'' {\em CoRR}, vol.~abs/2001.07761, 2020.

\bibitem{10074197}
A.~E. Ouadrhiri, A.~Abdelhadi, and P.~H. Phung, ``Hensel’s compression-based dimensionality reduction approach for privacy protection in federated learning,'' in {\em 2023 International Conference on Computing, Networking and Communications (ICNC)}, pp.~298--303, 2023.

\bibitem{10398775}
X.~Gao, ``Research on privacy protection scheme for educational data based on blockchain,'' in {\em 2023 3rd International Conference on Computer Science and Blockchain (CCSB)}, pp.~205--208, 2023.

\bibitem{Health:Privacy}
``Health information privacy.'' \url{https://www.hhs.gov/hipaa/for-professionals/privacy/index.html#:~:text=The%20HIPAA%20Privacy%20Rule%20establishes,care%20providers%20that%20conduct%20certain}, 2024.

\bibitem{HIPPA:Right}
``Individuals’ right under hipaa to access their health information.'' \url{https://www.hhs.gov/hipaa/for-professionals/privacy/guidance/access/index.html}, 2020.

\bibitem{kiya2023blockwiseencryptionreliablevision}
H.~Kiya, R.~Iijima, and T.~Nagamori, ``Block-wise encryption for reliable vision transformer models,'' 2023.

\bibitem{9802995}
Q.-X. Huang, W.~L. Yap, M.-Y. Chiu, and H.-M. Sun, ``Privacy-preserving deep learning with learnable image encryption on medical images,'' {\em IEEE Access}, vol.~10, pp.~66345--66355, 2022.

\bibitem{9767080}
B.~Shubyn, D.~Mrozek, L.~Fabry, T.~Maksymyuk, E.~M. Amhoud, and J.~Gazda, ``Federated learning techniques for 5g mobile networks,'' in {\em 2022 IEEE 16th International Conference on Advanced Trends in Radioelectronics, Telecommunications and Computer Engineering (TCSET)}, pp.~653--657, 2022.

\bibitem{9482536}
V.~A. Cunha, N.~Maroulis, C.~Papagianni, J.~Sacido, M.~Jiménez, F.~Ubaldi, M.~Gharbaoui, C.-Y. Chang, N.~Koursioumpas, K.~Tomakh, D.~Corujo, J.~P. Barraca, S.~Barmpounakis, D.~Kucherenko, A.~Giorgetti, A.~Boddi, L.~Valcarenghi, O.~Kolodiazhnyi, A.~Zabala, J.~X. Salvat, and A.~Garcia-Saavedra, ``5 growth: Secure and reliable network slicing for verticals,'' in {\em 2021 Joint European Conference on Networks and Communications \& 6G Summit (EuCNC/6G Summit)}, pp.~347--352, 2021.

\bibitem{wang2022protection}
Y.-R. Wang and Y.-C. Tsai, ``The protection of data sharing for privacy in financial vision,'' {\em Applied Sciences}, vol.~12, no.~15, p.~7408, 2022.

\bibitem{8359554}
C.~Huang, K.~Yan, S.~Wei, and D.~H. Lee, ``A privacy-preserving data sharing solution for mobile healthcare,'' in {\em 2017 International Conference on Progress in Informatics and Computing (PIC)}, pp.~260--265, 2017.

\bibitem{10461067}
V.~Mishra, K.~Gupta, D.~Saxena, and A.~K. Singh, ``A global medical data security and privacy preserving standards identification framework for electronic healthcare consumers,'' {\em IEEE Transactions on Consumer Electronics}, vol.~70, no.~1, pp.~4379--4387, 2024.

\bibitem{9153891}
A.~Qayyum, J.~Qadir, M.~Bilal, and A.~Al-Fuqaha, ``Secure and robust machine learning for healthcare: A survey,'' {\em IEEE Reviews in Biomedical Engineering}, vol.~14, pp.~156--180, 2021.

\bibitem{madono2020blockwisescrambledimagerecognition}
K.~Madono, M.~Tanaka, M.~Onishi, and T.~Ogawa, ``Block-wise scrambled image recognition using adaptation network,'' 2020.

\bibitem{8804201}
W.~Sirichotedumrong, T.~Maekawa, Y.~Kinoshita, and H.~Kiya, ``Privacy-preserving deep neural networks with pixel-based image encryption considering data augmentation in the encrypted domain,'' in {\em 2019 IEEE International Conference on Image Processing (ICIP)}, pp.~674--678, 2019.

\bibitem{dosovitskiy2020image}
A.~Dosovitskiy, L.~Beyer, A.~Kolesnikov, D.~Weissenborn, X.~Zhai, T.~Unterthiner, M.~Dehghani, M.~Minderer, G.~Heigold, S.~Gelly, {\em et~al.}, ``An image is worth 16x16 words: Transformers for image recognition at scale,'' {\em arXiv preprint arXiv:2010.11929}, 2020.

\bibitem{Lung}
``Lung and colon cancer histopathological images.'' \url{https://www.kaggle.com/datasets/andrewmvd/lung-and-colon-cancer-histopathological-images}, 2019.

\bibitem{msoud_nickparvar_2021}
M.~Nickparvar, ``Brain tumor mri dataset,'' 2021.

\end{thebibliography}
\end{document}